\documentclass[fleqn,12pt,twoside]{article}
\usepackage[headings]{espcrc1}

\readRCS
$Id: espcrc1.tex,v 1.2 2004/02/24 11:22:11 spepping Exp $
\usepackage{psfig}

\title{Soft Physics from STAR}

\author{Fuqiang Wang\address[purdue]{Department of Physics, Purdue University, West Lafayette, Indiana 47906, USA} (for the STAR Collaboration\thanks{For full list of STAR authors and acknowledgments, see appendix `Collaborations' of this volume.})
}
       
\runtitle{Soft Physics from STAR}
\runauthor{F. Wang}

\begin{document}

\maketitle

\begin{abstract}
New results on soft hadron distributions and correlations measured with the STAR experiment are presented. Knowledge about the bulk properties of relativistic heavy-ion collisions offered by these results is discussed.
\end{abstract}

\section{INTRODUCTION}

The mission of the Relativistic Heavy-Ion Collider (RHIC) is to create and study the Quark-Gluon Plasma (QGP), a {\em thermalized} system of {\em deconfined} quarks and gluons, long predicted to exist at high energy densities by Quantum Chromodynamics (QCD). Since the turn-on of RHIC in 2000, a large amount of data has been accumulated. STAR~\cite{cite1} and the other RHIC experiments~\cite{cite2} have critically assessed the first three years' wealth of data and concluded that a new form of hot and dense matter, with energy density far above the predicted critical energy density for hadron-QGP phase-transition~\cite{cite3}, has been created in heavy-ion collisions at RHIC.

Two new, successful runs, Run-IV and Run-V, were conducted. Results from these runs were presented at this Quark Matter conference. This proceedings presents an overview of soft physics results from STAR~\cite{cite4} and discusses what new knowledge, beyond that in~\cite{cite1}, can be learned from the results on the bulk properties of RHIC collisions. The overview focuses on three areas:
\begin{itemize}
\item[(1)] interactions between jets and the bulk medium, which probe the early stage of the medium, 
\item[(2)] elliptic flow, which is sensitive to the early stage equation of state, and 
\item[(3)] freeze-out bulk properties, which contain information accumulated over the evolution of the collision. 
\end{itemize}
For Hanbury Brown-Twiss (HBT) and small momentum correlations that are not covered in this overview the reader is referred to~\cite{cite5,cite6}, for event-by-event fluctuations to~\cite{cite7}, and for forward-rapidity physics to~\cite{cite8}.

\section{JET-MEDIUM INTERACTIONS}

Jets are produced early, by hard parton-parton scatterings. The scattered partons traverse the dense medium being created in heavy-ion collisions. They are coupled to the medium via strong interactions, lose energy, and fragment into hadrons, likely inside the medium. These fragment hadrons are predominantly soft, but possess characteristic jet-like angular correlations. The degree of energy loss and the changes in jet-like correlations depend on, and thus provide information on the gluon density of the medium~\cite{cite9}. The STAR detector with large acceptance, full azimuth coverage is ideal for jet correlation measurements. 
 
\subsection{Soft-Soft Correlations}

\begin{figure}[htb]
\vspace*{-0.2in}
\hspace*{0.1\textwidth}
\begin{minipage}{0.5\textwidth}
\psfig{file=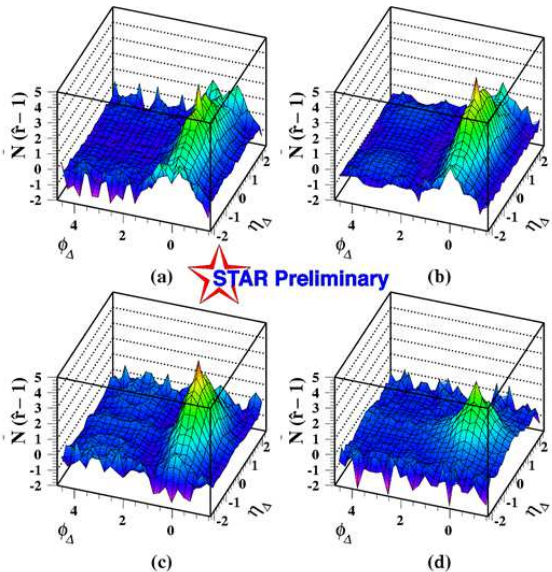,width=\textwidth}
\end{minipage}
\begin{minipage}{0.3\textwidth}
\caption{(color online) Angular correlations between soft particle pair of $0.15 < p_{\perp} < 2$~GeV/$c$ for (a) central to (d) peripheral collisions~\cite{cite11}. The $\eta_{\Delta}$   independent first and second harmonic terms in $\phi_{\Delta}$  have been subtracted.\label{fig1}}
\end{minipage}
\end{figure}

STAR has measured angular correlations between two low $p_{\perp}$ particles without coincidence requirement with a high $p_{\perp}$ particle. Clear jet correlation structures, even at $p_{\perp}$ as low as 0.6-0.8~GeV/$c$, are observed in p+p collisions~\cite{cite10}. Such correlation analysis is also carried out in heavy-ion collisions. Figure~\ref{fig1} shows angular correlations in 130~GeV Au+Au collisions between two soft particles of $0.15<p_{\perp}<2$~GeV/$c$~\cite{cite11}. The first and second harmonic terms have been subtracted. The small-angle correlation peak, characteristic of (mini-)jets, narrows in $\phi$  with centrality, and more dramatically, broadens in $\eta$ by a factor of 2.3 from peripheral to central collisions~\cite{cite11}. The results demonstrate the strong coupling between these correlated particles and the medium.

\vspace*{-0.3in}
\subsection{Hard-Soft Correlations}

Jets can be more cleanly selected by triggering on high $p_{\perp}$ particles; A high $p_{\perp}$ particle likely selects di-jets from hard parton-parton scatterings. The hard-scattered partons lose energy in the medium, emerging as lower $p_{\perp}$ particles than expected from fragmentation in vacuum. Due to energy loss, the measured high $p_{\perp}$ particles come preferentially from jets produced on the surface of the collision zone and directed outward. The partner jet, directed inward, suffers maximal energy loss resulting in the observed depletion of high $p_{\perp}$ particles and enhancement of low $p_{\perp}$ particles~\cite{cite12,cite13}. The low $p_{\perp}$ particles are broadly distributed and appear not much harder than the inclusive hadrons from medium decay~\cite{cite13}. This illustrates, experimentally, the thermalization processes in heavy ion-collisions: particles from two distinctively different sources, jets and medium, approach equilibration via parton-parton interactions. 

\begin{figure}[htb]
\hspace*{0.1\textwidth}
\begin{minipage}{0.355\textwidth}
\psfig{file=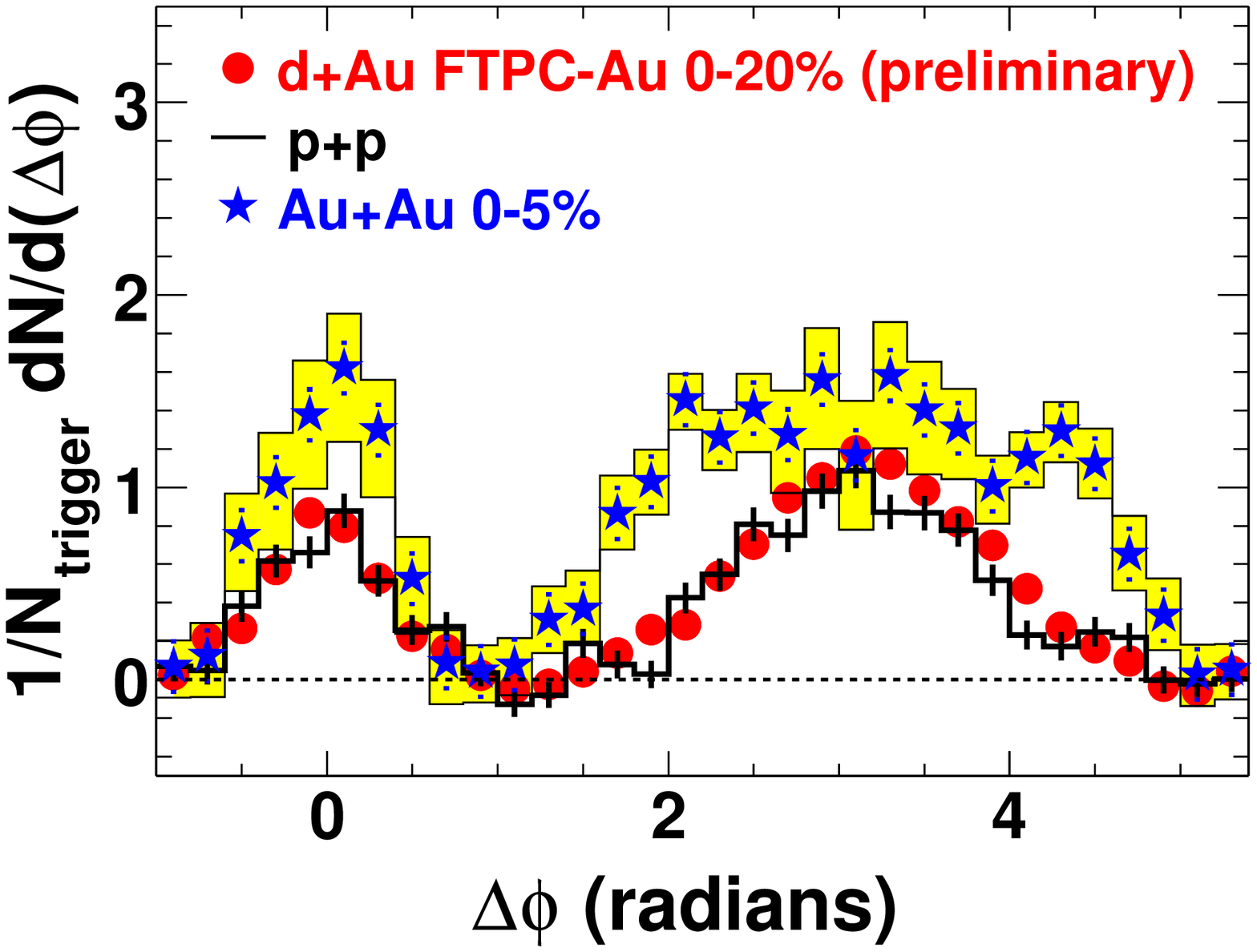,width=\textwidth}
\vspace*{-0.34in}
\psfig{file=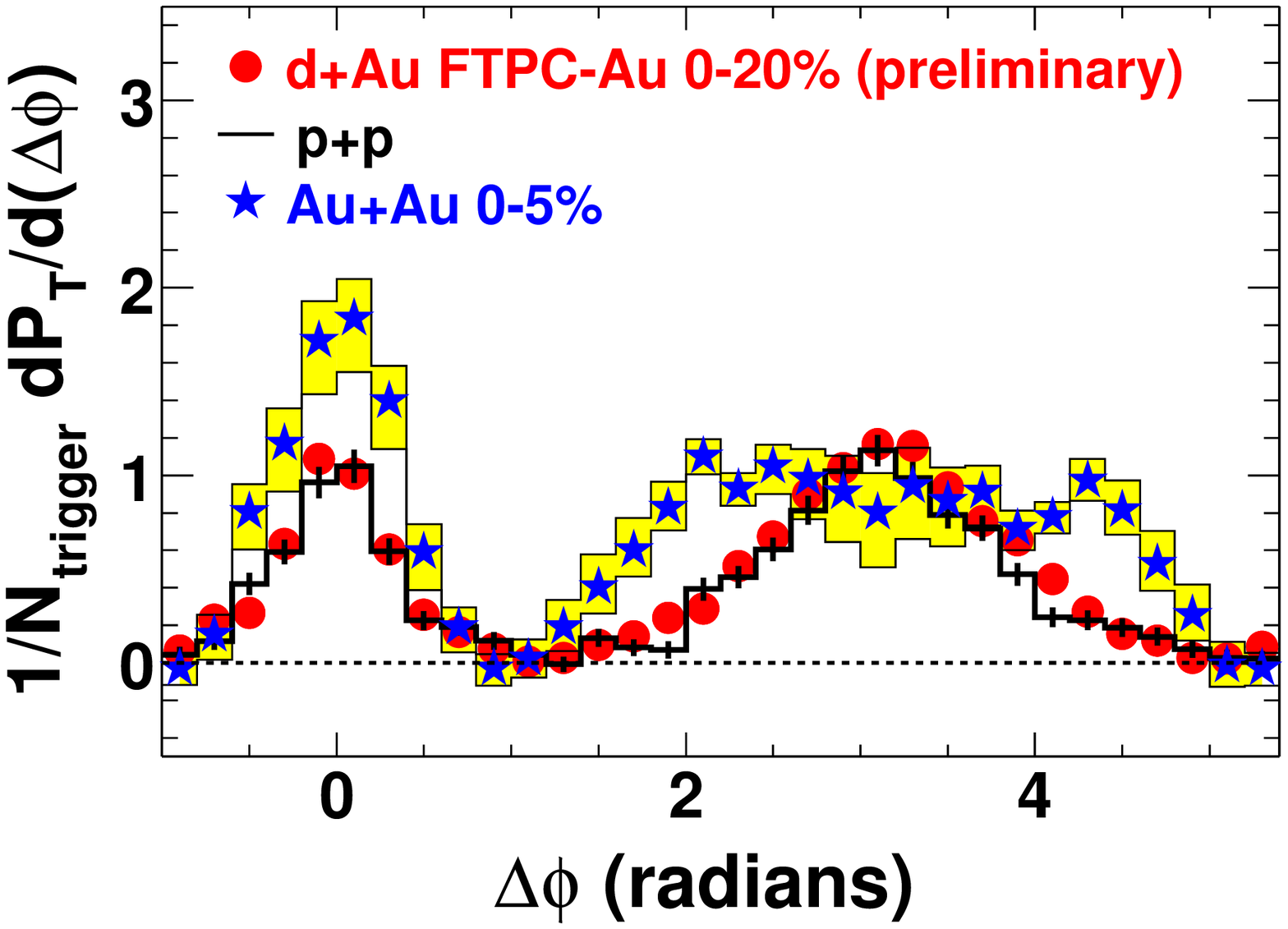,width=\textwidth}
\end{minipage}
\begin{minipage}{0.445\textwidth}
\vspace*{-0.2in}
\psfig{file=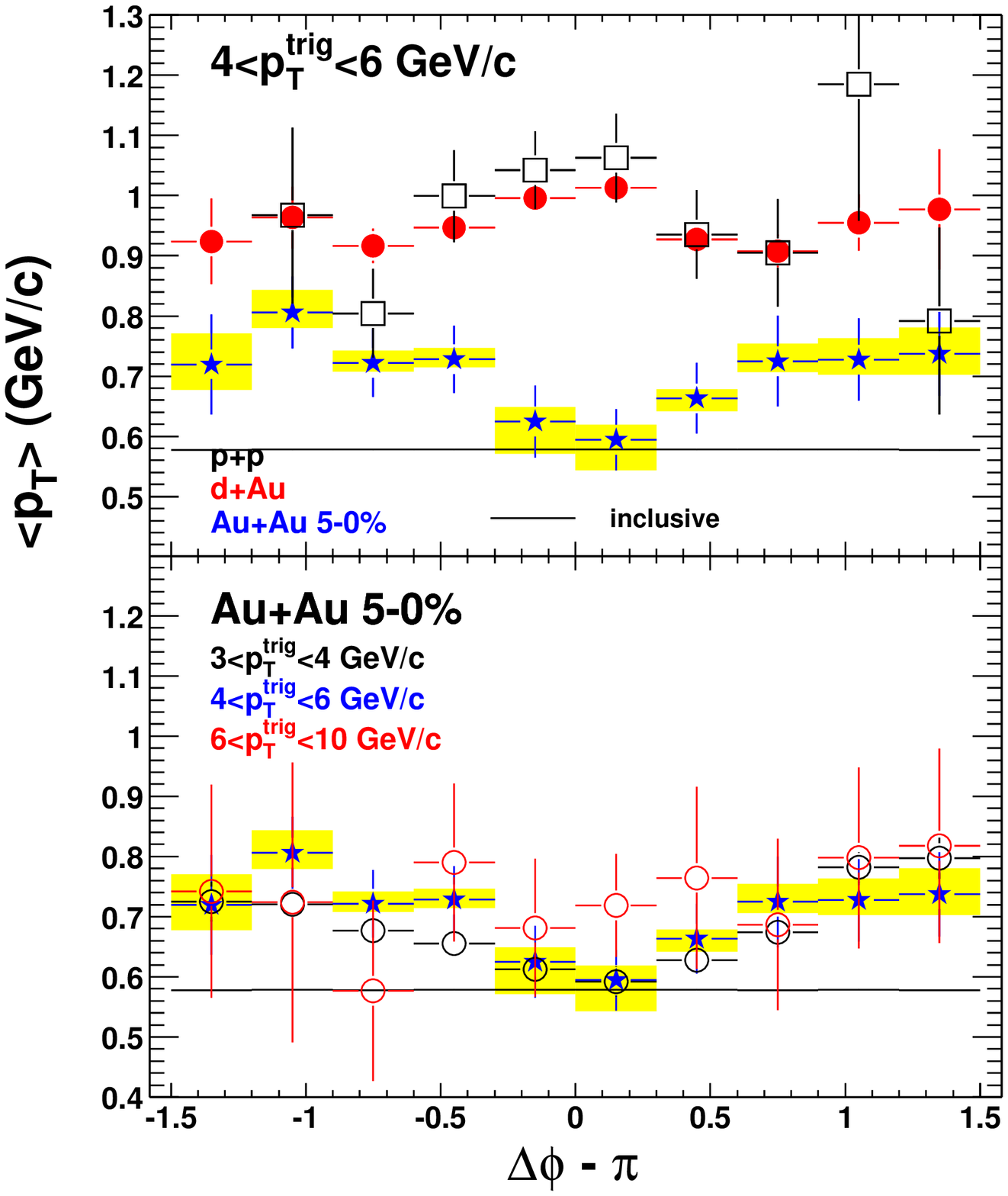,width=\textwidth}
\end{minipage}
\vspace*{-0.25in}
\caption{(color online) Left: Background subtracted number (upper) and $p_{\perp}$-weighted (lower) correlation functions in $pp$, central 20\% d+Au and 5\% Au+Au collisions~\cite{cite14}. Right: The $\langle p_{\perp} \rangle$ of associated hadrons on the away side for the three systems (upper) and three trigger $p_{\perp}$ selections (lower). The shaded areas are systematic uncertainties.}
\label{fig2}
\end{figure}

\vspace*{-0.3in}
STAR has further studied the azimuthal dependence of associated $\langle p_{\perp} \rangle$. Figure~\ref{fig2}(left) shows the number and $p_{\perp}$-weighted correlation functions in $pp$, d+Au, and central Au+Au collisions~\cite{cite14,cite15}. The subtracted background is obtained from mixed-events, modulated by elliptic flow and normalized to the signal in the $0.8<| \Delta\phi |<1.2$ region, and is the major source of systematic uncertainties~\cite{cite13}. The $pp$ and d+Au data are similar, and the Au+Au correlation is significantly broader. Figure~\ref{fig2}(right) shows the $\langle p_{\perp} \rangle$, obtained from the ratio of the correlation functions, as a function of $\Delta\phi$    on the away side. The inclusive hadron $\langle p_{\perp} \rangle$ is shown as line. The $\langle p_{\perp} \rangle$ for $pp$ and d+Au are peaked at $\Delta\phi  = \pi$, as expected from jet fragmentation, and are much larger than the inclusive one. For central Au+Au collisions, however, the $\langle p_{\perp} \rangle$ is smallest at  $\Delta\phi  = \pi$  for the two low trigger $p_{\perp}$ selections and appears to be equal to the inclusive $\langle p_{\perp} \rangle$, while at other angles it is between the values for the simpler systems (pp and d+Au) and the inclusive data. 

The $\langle p_{\perp} \rangle$ versus $\Delta\phi$    result indicates that the spectrum at  $\Delta\phi=\pi$    is softer than that further away from  $\pi$. In other words, relatively fewer high $p_{\perp}$ associated particles are found at  $\pi$. This is also demonstrated in the correlation functions with varying associated $p_{\perp}$: with increasing associated $p_{\perp}$, the correlation function flattens and even develops double-hump structure~\cite{cite14,cite16}. 

\subsection{Three-Particle Correlations}

Our correlation function results are qualitatively consistent with a number of scenarios. In one scenario, the away-side jets directed at $\pi$  are maximally modified due to the maximum path-length through the medium. In an alternate scenario, the measured away-side correlation is the sum of away-side jets deflected by radial flow, each individual jet relatively confined in azimuth. Because the jet axis is now deflected at an angle away from $\pi$  and high $p_{\perp}$ jet fragments are relatively confined to the jet axis, a dip at $\pi$  is resultant in the correlation function at high associated $p_{\perp}$. In a third scenario, the energy deposited by the away-side jet generates a sonic shock wave in the medium, producing Mach cone effect: larger number of particles and possibly larger $\langle p_{\perp} \rangle$ along the conical flow direction, approximately at $\Delta\phi  = \pi \pm 1$~\cite{cite17}. 

\begin{figure}[htb]
\vspace*{-0.4in}
\centerline{\psfig{file=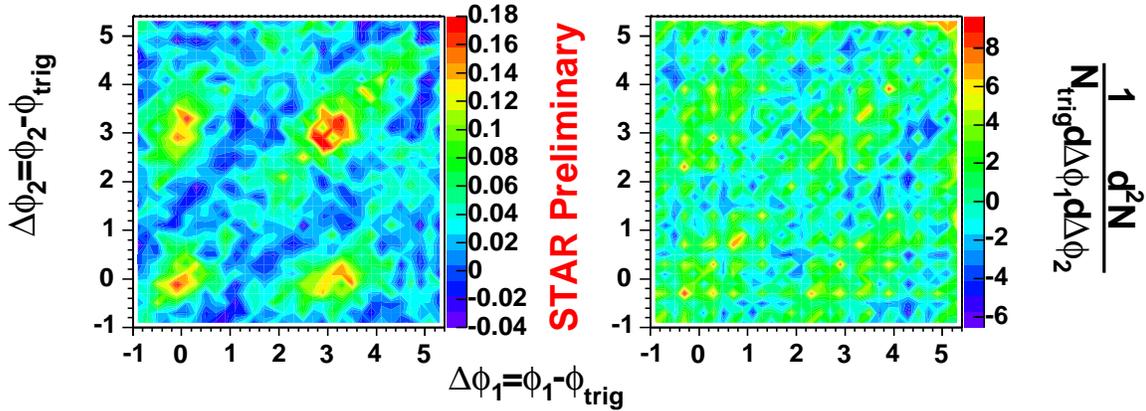,width=\textwidth}}
\vspace*{-0.3in}
\caption{(color online) Three-particle correlations in min-bias d+Au and 10\% central Au+Au collisions~\cite{cite14}. The $p_{\perp}$ ranges are $p_{\perp}^{\rm trig}$=3-4~GeV/$c$ for the trigger and $p_{\perp}$=1-2~GeV/$c$ for the associated particles.}
\label{fig3}
\end{figure}

In order to distinguish various scenarios, 3-particle correlations are analyzed. Figure~\ref{fig3} shows 3-particle correlation results in minimum bias d+Au and 10\% central Au+Au collisions for trigger $p_{\perp}$=3-4~GeV/$c$ and associated $p_{\perp}$=1-2~GeV/$c$~\cite{cite14}. A number of combinatoric backgrounds have been subtracted: 
\begin{itemize}
\item[(a)] Correlated trigger-associated pair (see section 2.2) with a background particle. This background is constructed by coupling the 2-particle correlation function measured in the same event with the $v_{2}$ modulated background.
\item[(b)] Correlated soft-soft hadron pair (see section 2.1) in the underlying event that is not related to the trigger particle. This background is constructed from soft-soft correlations in non-triggered events.
\item[(c)] Random triplets that are correlated by elliptic flow. This background is constructed by $1+2v_{2}^{\rm trig}v_{2}^{(1)}\cos(\Delta\phi_1)+2v_{2}^{\rm trig}v_{2}^{(2)}\cos(\Delta\phi_2)+2v_{2}^{(1)}v_{2}^{(2)}\cos(\Delta\phi_1-\Delta\phi_2)$ and normalized to the signal in the region of $0.8<|\Delta\phi_{1,2}|<1.2$.
\end{itemize}

\vspace*{-0.2in}
As seen in Fig.~\ref{fig3}, four peaks are evident in both d+Au and Au+Au, corresponding to near-near, near-away, away-near, and away-away correlations for the two associated particles. The away-away peak is elongated along the diagonal axis, in d+Au perhaps due to $k_{\perp}$ broadening, and in Au+Au with possible additional effect from deflected jets. The correlation strength is stronger in Au+Au than in d+Au and is qualitatively consistent with the relative two-particle correlation strengths. 

Conical flow would yield 3-particle correlation signals at ($\pi \pm 1, \pi \mp 1$) along the off-diagonal axis. To investigate this possibility, we take the differences between the averaged correlation signals in three regions, `{\it center}': $|\Delta\phi_{1,2} - \pi|<0.4$, `{\it deflected}': $|\Delta\phi_{1,2}-(\pi \pm 1)|<0.4$, and `{\it cone}': $|\Delta\phi_1-(\pi \pm 1)|<0.4$ and $|\Delta\phi_2-(\pi \mp 1)|<0.4$. The differences in 10\% central Au+Au collisions per radian$^{2}$ are $center - deflected = 0.3 \pm 0.3{\rm (stat)} \pm 0.4{\rm (syst)}$, and $center - cone = 2.6 \pm 0.3{\rm (stat)} \pm 0.8{\rm (syst)}$. The systematic uncertainties are obtained from those in $v_{2}$, taken to be between 4-particle cumulant and reaction plane results, and those in background normalization. The latter affects greatly the absolute correlation magnitudes, but not the differences. The results indicate that the correlation strength is significantly lower in the conical flow `cone' area than in the `center' or `deflected' area; The distinctive features of conical flow are not observed in the present data.

\section{ELLIPTIC FLOW}

In non-central collisions, the overlap region is anisotropic (nearly elliptic). Large pressure built up in the collision center results in pressure gradient dependent on azimuthal angle, which generates momentum space anisotropy, or elliptic flow. Once the spatial anisotropy disappears due to the anisotropic expansion, development of elliptic flow ceases. This self-quenching process happens quickly, therefore elliptic flow is primarily sensitive to the early stage equation of state (EOS)~\cite{cite18}.

\subsection{Hydrodynamic Description}

\begin{figure}[htb]
\begin{minipage}{0.63\textwidth}
\vspace*{-0.35in}
\psfig{file=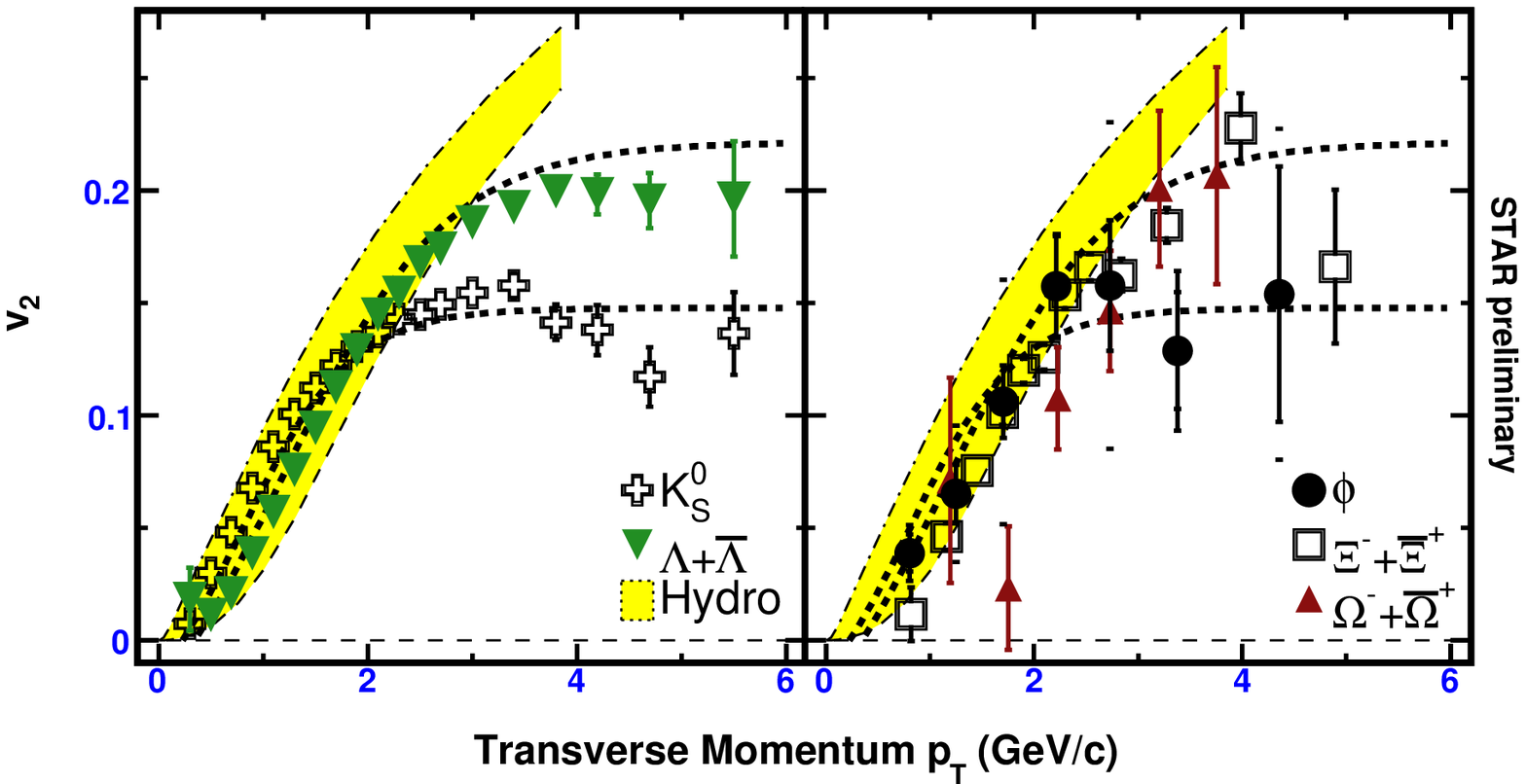,width=\textwidth}
\end{minipage}
\begin{minipage}{0.34\textwidth}
\vspace*{-0.7in}
\caption{(color online) Azimuthal anisotropy $v_{2}$ for strange (left) and (hidden) multi-strange (right) hadrons~\cite{cite19}. The curves are empirical fits. The shaded areas are ranges of hydro results.}
\label{fig4}
\end{minipage}
\begin{minipage}{0.63\textwidth}
\vspace*{-0.2in}
\centerline{\psfig{file=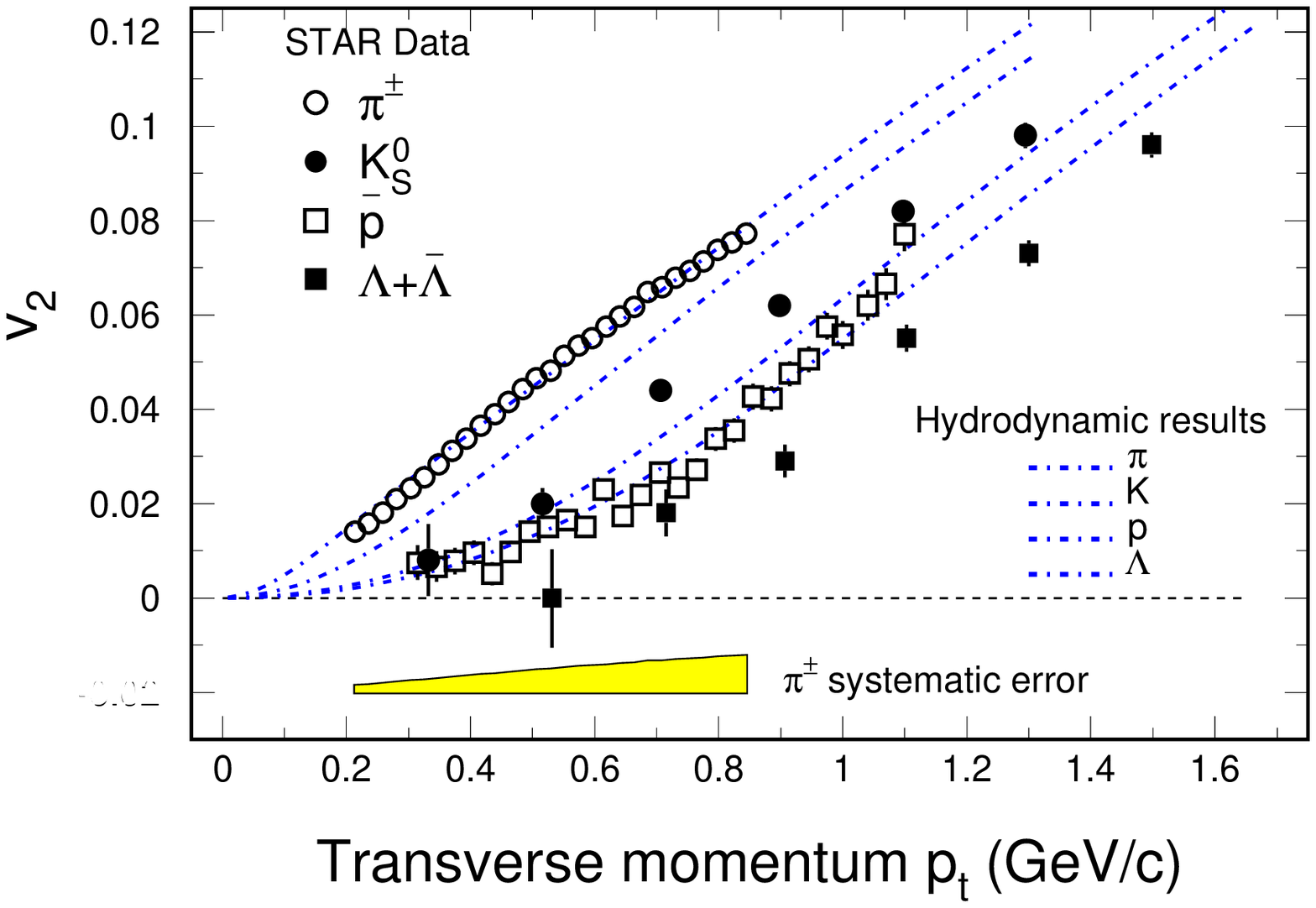,width=0.9\textwidth}}
\end{minipage}
\begin{minipage}{0.34\textwidth}
\vspace*{-0.2in}
\caption{(color online) Elliptic flow $v_{2}$ as a function of  $p_{\perp}$ from minimum bias Au+Au collisions~\cite{cite21}. Hydrodynamic calculations are shown in dot-dashed lines.}
\label{fig5}
\end{minipage}
\end{figure}

Figure~\ref{fig4} shows the measured elliptic flow parameter, $v_{2}$, as a function of $p_{\perp}$ for $K^0_S$, $\Lambda$, $\phi$, $\Xi$ and $\Omega$ ~\cite{cite19,cite20}. Large $v_{2}$ is observed for all particle species, indicating strong interactions at the early stage. Since $\phi$, $\Xi$ and $\Omega$  have low hadronic cross-sections, their observed large $v_{2}$ may suggest that the elliptic flow is built up in the partonic stage. 

\vspace*{-0.43in}
Figure~\ref{fig4} also shows the range of $v_{2}$ from hydrodynamic calculations. A more detailed comparison is shown in Fig.~\ref{fig5} where the hydro results, using an EOS with a first-order hadron-QGP phase-transition~\cite{cite18}, are compared to pion, kaon, proton, and lambda $v_{2}$ measurements~\cite{cite21}. The mass dependence observed in data, characteristic of a common flow velocity, is well described by hydrodynamics. The magnitude of the $v_{2}$, however, is described only at the 20-30\% level. The description with hadronic EOS is worse~\cite{cite18}. The discrepancy indicates perhaps that the hydro calculation is too simplistic and more realistic EOS is needed. On the other hand, non-flow effects in the measured $v_{2}$ need to be addressed rigorously. Nevertheless, since ideal hydrodynamic fluid is a thermalized system with zero mean-free-path yielding the maximum possible $v_{2}$, the approximate consistency between the measured $v_{2}$ and the hydro results suggests an early thermalization in heavy-ion collisions at RHIC.

\subsection{Constituent Quark Scaling}

\begin{figure}[htb]
\hspace*{0.1\textwidth}
\begin{minipage}{0.5\textwidth}
\vspace*{-0.35in}
\centerline{\psfig{file=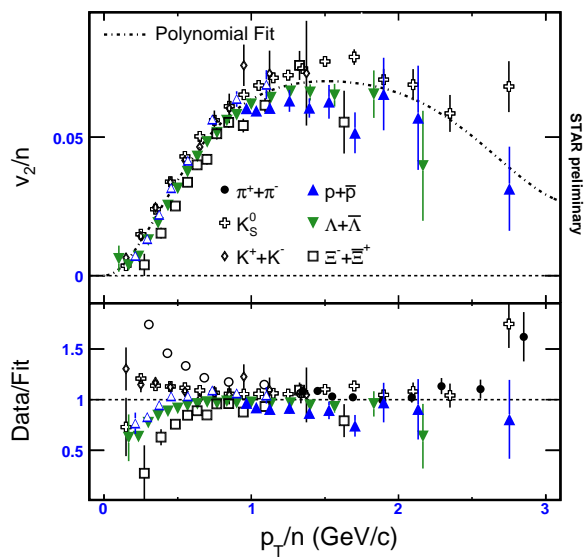,width=\textwidth}}
\end{minipage}
\begin{minipage}{0.35\textwidth}
\caption{(color online) Measurements of scaled $v_{2}(p_{\perp}/n)/n$ for identified hadrons and ratios between the measurements and a polynomial fit through all data points~\cite{cite19}.}
\end{minipage}
\label{fig6}
\end{figure}

While well described by hydrodynamics at low $p_{\perp}$, $v_{2}$ was found to saturate at high $p_{\perp} > 2$~GeV/$c$~\cite{cite19}. The saturation value for mesons is about 2/3 of that for baryons. This separation pattern holds for $\pi$, $K$, $p$, $\Lambda$, and $\Xi$, and seems to hold for $\phi$  and $\Omega$~\cite{cite20,cite22}. This result, together with the baryon-meson splitting of high $p_{\perp}$ suppression pattern~\cite{cite23}, suggests the relevance of the constituent quark degrees of freedom in the intermediate $p_{\perp}$ region~\cite{cite24}. Figure~\ref{fig6} shows new, high precision measurement of $v_{2}$, scaled by the number of valence quarks $n$, as a function of $p_{\perp}$ also scaled by $n$~\cite{cite19}. Although deviation is visible, constituent quark scaling still holds to a good degree.

\vspace*{-0.35in}
{\em If} constituent quark coalescence is indeed the production mechanism giving rise to the baryon-meson difference at intermediate $p_{\perp}$, i.e. hadronization of the bulk medium does occur, then it seems natural to conclude that a {\em deconfined} phase of quarks and gluons is created, prior to hadronization.

\subsection{Angular Correlations with Baryons and Mesons}

\begin{figure}[htb]
\vspace*{-0.3in}
\centerline{\psfig{file=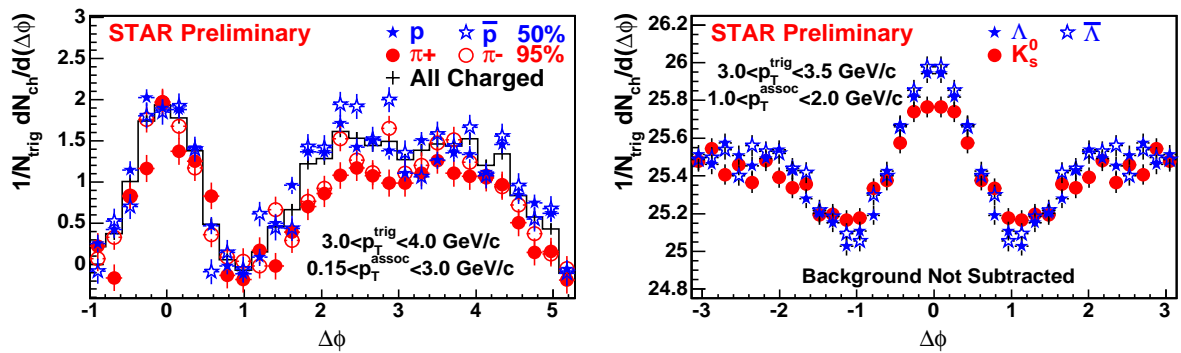,width=\textwidth}}
\vspace*{-0.35in}
\caption{(color online) Azimuthal correlation functions of charged hadrons with identified trigger particles, $\Lambda$  and $K_S$ ($p_{\perp}^{\rm trig}$=3-3.5~GeV/$c$, $p_{\perp}$=1-2~GeV/$c$) and $p$, $\bar{p}$, $\pi^+$  and $\pi^-$     ($p_{\perp}^{\rm trig}$=3-4~GeV/$c$, $p_{\perp}$=0.15-3~GeV/$c$)~\cite{cite14}.}
\label{fig7}
\end{figure}

Constituent quark coalescence/recombination~\cite{cite25} is expected to produce either no jet-like angular correlations, or weaker ones than those from jet fragmentation. Figure~\ref{fig7} shows correlation results for charged hadrons with $\Lambda$, $K_S$, $p$, or $\pi$  of $3 < p_{\perp}^{\rm trig} < 4$~GeV/$c$~\cite{cite14}. The $\Lambda$  and $K_S$ are identified by their topological decays in the TPC; the $p$ and $\pi$ at high $p_{\perp}$ are statistically identified by the relativistic rise of the specific ionization energy loss in the TPC and are required to have 50\% and 95\% purity, respectively. Little difference is found in Fig.~\ref{fig7} between baryon and meson triggers, or particle and anti-particle triggers. The lack of difference in jet-like correlations between leading baryons and mesons and the success of coalescence and recombination models in describing intermediate $p_{\perp}$ particle yields and $v_{2}$ are hard to reconcile, but may provide a unique window to advance our understanding of hadronization in heavy-ion collisions.

\vspace*{-0.4in}
\section{FREEZE-OUT BULK PROPERTIES}

If thermalization is reached at an initial stage, the final stage hadrons will possess thermal distributions. Indeed, the measured particle spectra and yields~\cite{cite26} and event-by-event $\langle p_{\perp} \rangle$ fluctuations~\cite{cite27} indicate a nearly chemically and (local-)kinetically equilibrated system at the final freeze-out stage. 

\subsection{Chemical and Kinetic Freeze-out Parameters}

\begin{figure}[htb]
\vspace*{-0.2in}
\centerline{
\psfig{file=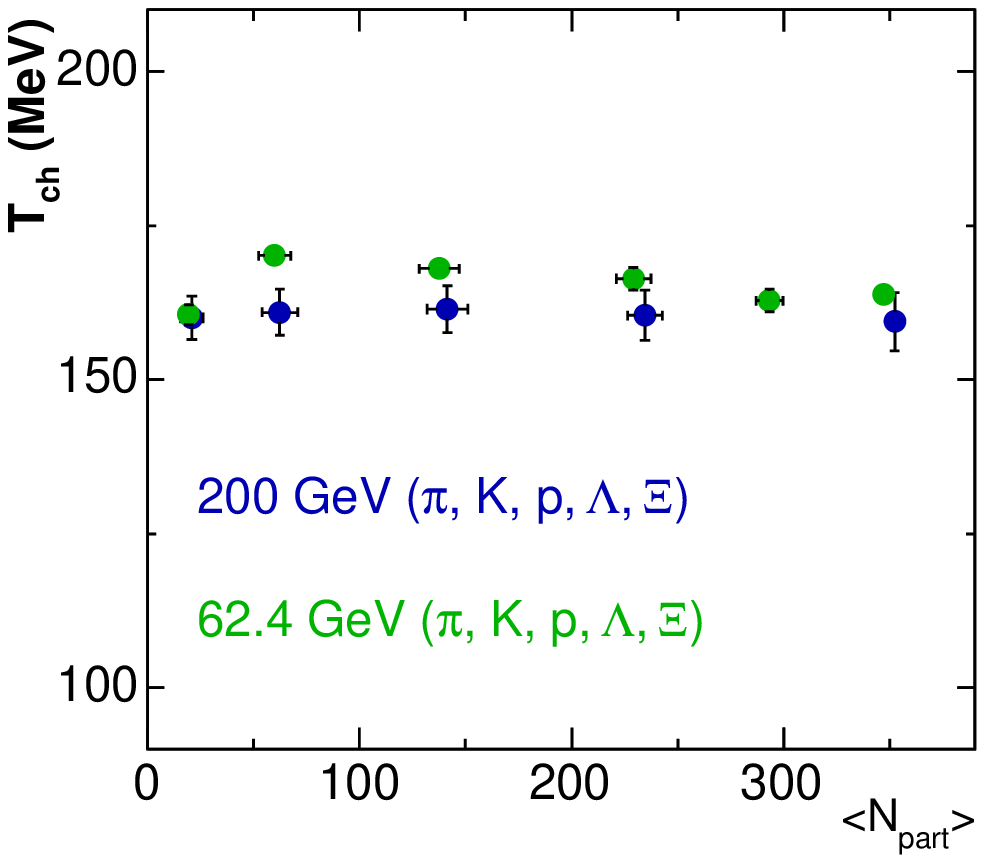,width=0.4\textwidth,height=0.3\textwidth}
\psfig{file=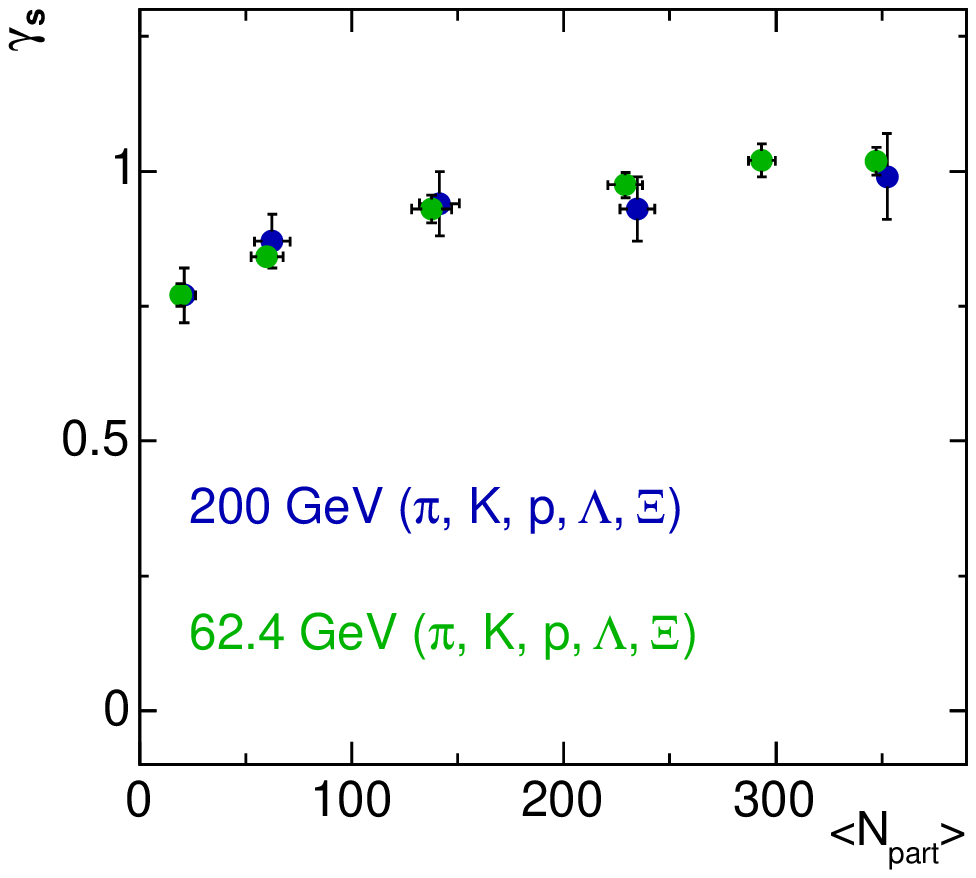,width=0.4\textwidth,height=0.3\textwidth}
}
\vspace*{-0.35in}
\caption{(color online) Extracted chemical freeze-out temperature (left) and strangeness suppression factor (right) from stable particle ratios~\cite{cite28,cite29}.}
\label{fig8}
\vspace*{0.2in}
\begin{minipage}{0.49\textwidth}
\centerline{\psfig{file=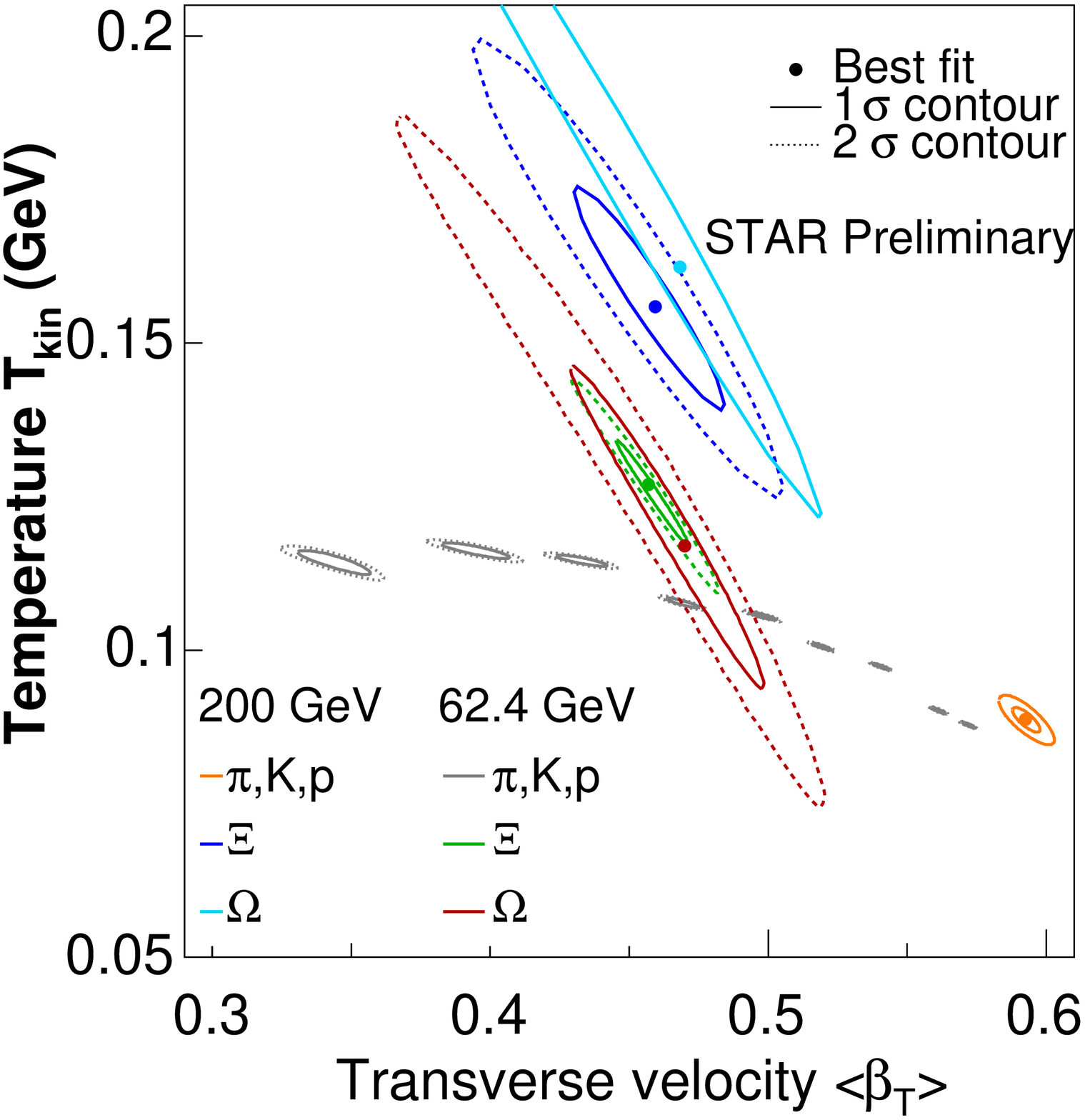,width=0.92\textwidth,height=0.90\textwidth}}
\vspace*{-0.3in}
\caption{(color online) Extracted kinetic freeze-out temperature versus average flow velocity within the blast-wave model~\cite{cite28,cite29}.}
\label{fig9}
\end{minipage}
\hspace*{0.01\textwidth}
\begin{minipage}{0.49\textwidth}
\psfig{file=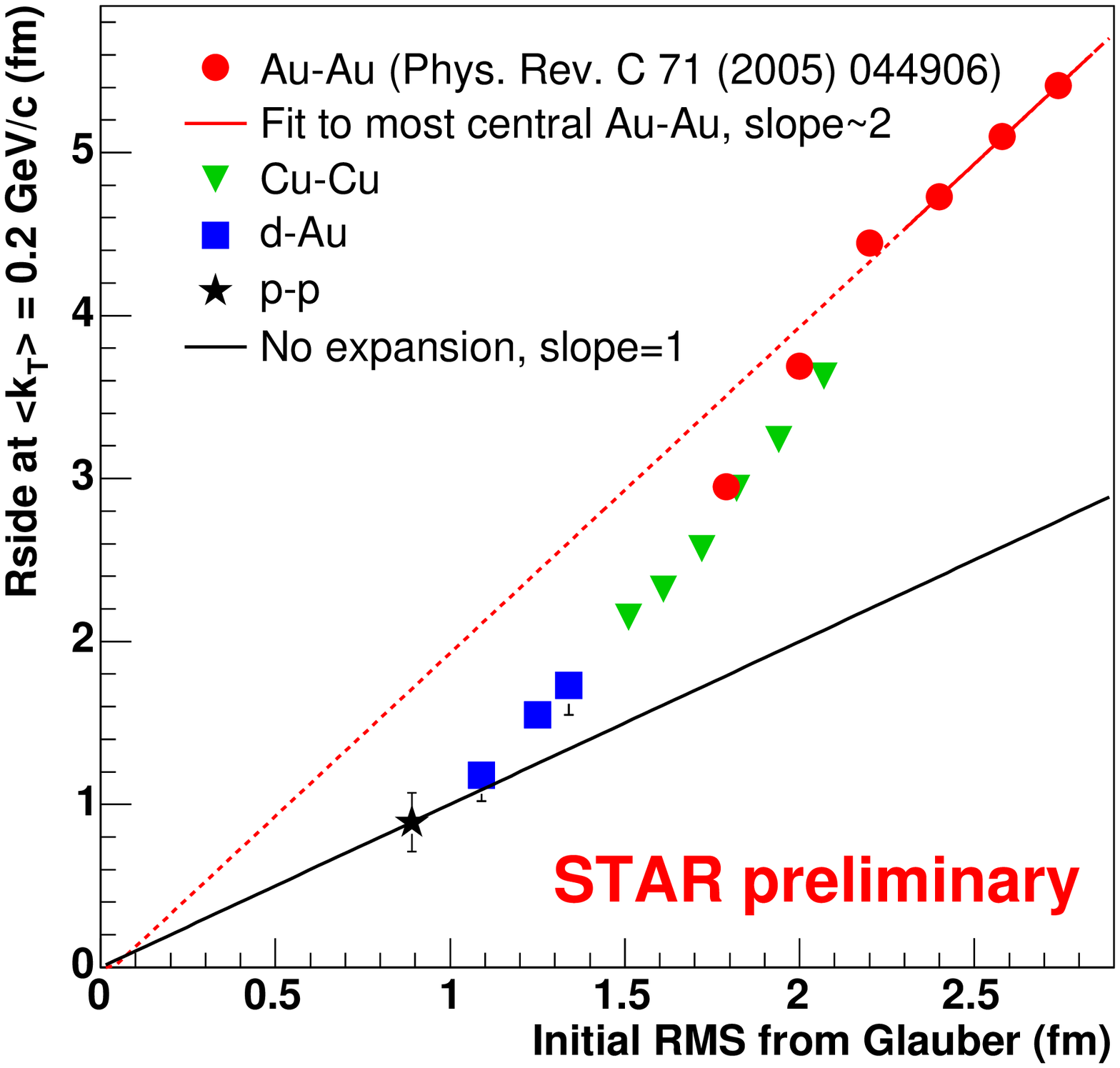,width=0.93\textwidth}
\vspace*{-0.37in}
\caption{(color online) Sideward size from $2\pi$  HBT measurements versus initial RMS size of the collision system calculated by Glauber model~\cite{cite5}.}
\label{fig10}
\end{minipage}
\end{figure}

STAR~\cite{cite28,cite29} has now measured hadron distributions at 62~GeV, and extracted chemical freeze-out properties from stable particle ratios within the thermal model~\cite{cite30} and kinetic freeze-out properties from particle $p_{\perp}$ distributions within the blast-wave model~\cite{cite31}. Figure~\ref{fig8} shows the extracted chemical freeze-out temperature ($T_{\rm CH}$) and strangeness suppression factor ($\gamma_S$), and Fig.~\ref{fig9} the extracted kinetic freeze-out temperature ($T_{\rm KIN}$) and average radial flow velocity ($\langle\beta_{\perp}\rangle$). Resonance decays are found to have no significant effect on the extracted kinetic freeze-out parameters~\cite{cite28}. The results at 62~GeV are qualitatively the same as those obtained at 200~GeV. 

It is found that the extracted $T_{\rm CH}$ is independent of centrality and is close to the predicted hadronization temperature~\cite{cite3}. The $T_{\rm KIN}$, extracted from distributions of {\em common} particle species ($\pi$, $K$, $p$), is found to steadily decrease with centrality, while the corresponding $\langle\beta_{\perp}\rangle$ increases~\cite{cite26,cite32}. This provides evidence for further expansion from chemical to kinetic freeze-out, driving the system to lower temperature. Thus one may argue that the constant $T_{\rm CH}$, {\em alone}, suggests that chemical freeze-out measures hadronization; hadronic scatterings from hadronization (at a universal temperature) to chemical freeze-out must be negligible because they would result in a dropping $T_{\rm CH}$. 

The extracted $\gamma_S$ is found to increase from $pp$ and peripheral collisions to central collisions. In central Au+Au collisions, $\gamma_S$ is found to be consistent with unity, suggesting that strangeness is similarly equilibrated as the light u,d quarks.
The extracted $T_{\rm KIN}$ for {\em rare} particles ($\phi$, $\Xi$, $\Omega$) is higher than that for the common particles and appears to be independent of centrality and close to $T_{\rm CH}$. The flow velocity is lower, but sizeable. This may indicate, especially noting their low hadronic cross-sections, that these rare particles kinetically and chemically freeze-out at the same time; the measured  $\langle\beta_{\perp}\rangle$  for the rare particles then can be considered as the flow velocity at chemical freeze-out (and perhaps also at hadronization). This is in agreement with the $v_{2}$ results: a significant amount of flow is built up at the partonic stage.

\vspace*{-0.3in}
\subsection{System Size and Time Span}

It is found that resonance yields are lower than predicted by the thermal model using the fitted parameters to stable particle ratios~\cite{cite33}. This implies a finite time span from chemical to kinetic freeze-out during which resonances decay and the daughter particles rescatter resulting in a loss of resonance signals. On the other hand, regeneration of resonances from particle rescattering is also possible. Thus, yields of short-lived resonances are determined by conditions after chemical freeze-out, and effectively measure the kinetic freeze-out temperature.

System size and time information can be measured by HBT. Figure~\ref{fig10} shows the sideward size ($R_{side}$) of the freeze-out system measured by HBT of pion pairs at low $k_{\perp}$~\cite{cite5}. Due to radial flow, the true $R_{side}$, to be found at $k_{\perp}=0$, is about 20\% larger than these in Fig.~\ref{fig10}. $R_{side}$ is equal to the initial RMS size in $pp$ and peripheral d+Au and becomes twice as large in central Au+Au collisions, with a smooth trend in-between. This result indicates an expansion of a factor of 2 from initial to final state in central Au+Au. There remain a number of interesting questions: What is the size at chemical freeze-out? Will it, together with flow velocity, be consistent with the resonance results? Will hydrodynamic-like models, successful for $v_{2}$ and hadron spectra results, be able to explain HBT results?

\section{CONCLUSIONS}

STAR has measured wealth of data pertinent to properties of the bulk medium created in relativistic heavy-ion collisions. The new results from Run-IV and Run-V, some of which are reviewed here, have furthered our understanding and yielded new insights. We summarize the results as follow. 
\begin{itemize}
\item[(1)] Jet correlations are strongly modified by the medium. Correlated hadrons are nearly equilibrated with the medium via strong interactions of partons, or jets, or both with the medium. The distinctive features of conical flow are not observed in our present three-particle correlation data. 
\item[(2)] Elliptic flow is large, even for rare particles such as $\phi$,  $\Xi$, and  $\Omega$. Elliptic flow can be reasonably well described by hydrodynamics at low $p_{\perp}$, and at high $p_{\perp}$ displays saturation patterns grouping into baryons and mesons. The results suggest partonic collectivity, early thermalization, and the relevance of the constituent quark degrees of freedom. Quark coalescence offers a simple mechanism for hadronization; however, it appears at odd with the jet-like correlation results with trigger baryons and mesons. 
\item[(3)] Extracted chemical freeze-out temperature does not change with centrality; chemical freeze-out may coincide with hadronization with significant radial flow. Pion, kaons, protons favor a kinetic freeze-out changing with centrality, and together with resonance yields, suggest a finite time span from chemical to kinetic freeze-out. 
\end{itemize}


\begin{thebibliography}{99}
\bibitem{cite1}  J. Adams {\it et al.} (STAR Collaboration), Nucl. Phys. {\bf A757}, 102-183 (2005).
\bibitem{cite2}  K. Adcox {\it et al.} (PHENIX Collaboration) Nucl. Phys. {\bf A757}, 184-283 (2005);  I.	Arsene {\it et al.} (BRAHMS Collaboration) {\it ibid}, {\bf A757}, 1-27 (2005); B.B. Back {\it et al.} (PHOBOS Collaboration) {\it ibid}, {\bf A757}, 28-101 (2005).
\bibitem{cite3}  F. Karsch, Nucl Phys {\bf A698}, 199c (2002).
\bibitem{cite4}  M. Anderson {\it et al.}, Nucl. Instrum. Meth. {\bf A499}, 659 (2003); K.H.Ackermann {\it et al.}, {\it ibid}, {\bf A499}, 713 (2003).
\bibitem{cite5}  Z. Chajecki {\it et al.} (STAR Collaboration),  these proceedings.
\bibitem{cite6}  P. Chaloupka {\it et al.} (STAR Collaboration), these proceedings.
\bibitem{cite7}  C. Pruneau {\it et al.} (STAR Collaboration), these proceedings.
\bibitem{cite8}  B. Mohanty {\it et al.} (STAR Collaboration), these proceedings.
\bibitem{cite9}  X.-N. Wang and M. Gyulassy, Phys Rev Lett {\bf 68}, 1480 (1992); R. Baier, D. Schiff, and B.G. Zakharov, Ann Rev Nucl Part Sci {\bf 50}, 37 (2000).
\bibitem{cite10}  T. Trainor {\it et al.} (STAR Collaboration), poster presented at this conference.
\bibitem{cite11}  J. Adams, {\it et al.} (STAR Collaboration), nucl-ex/0411031.
\bibitem{cite12}  C. Adler {\it et al.} (STAR Collaboration), Phys. Rev. Lett. {\bf 90}, 082302 (2003).
\bibitem{cite13}  J. Adams {\it et al.} (STAR Collaboration), Phys. Rev. Lett. {\bf 95}, 152301 (2005) [nucl-ex/0501016]; F. Wang (STAR Collaboration), J. Phys. G {\bf 30}, S1299 (2004).
\bibitem{cite14}  J. Ulery {\it et al.} (STAR Collaboration), these proceedings.
\bibitem{cite15}  F. Wang {\it et al.} (STAR Collaboration), nucl-ex/0508021.
\bibitem{cite16}  M. Horner {\it et al.} (STAR Collaboration), poster presented at this conference.
\bibitem{cite17}  H. Stoecker, nucl-th/0406018; J. Casalderrey-Solana, E.V. Shuryak, and D. Teaney,  hep-ph/0411315; H. Stoecker, nucl-th/0506013.
\bibitem{cite18}  P. Huovinen, nucl-th/0305064; P. Kolb and U. Heinz, nucl-th/0305084; E.V. Shuryak,  Prog Part Nucl Phys {\bf 53}, 273 (2004).
\bibitem{cite19}  M. Oldenburg {\it et al.} (STAR Collaboration), these proceedings.
\bibitem{cite20}  X. Cai {\it et al.} (STAR Collaboration), these proceedings.
\bibitem{cite21}  J. Adams {\it et al.} (STAR Collaboration), Phys. Rev. C {\bf 72}, 014904 (2005).
\bibitem{cite22}  J. Adams {\it et al.} (STAR Collaboration), Phys Rev Lett {\bf 92}, 052302 (2004); J. Adams {\it et al.} (STAR Collaboraiton) nucl-ex/0504022.
\bibitem{cite23}  J. Dunlop {\it et al.} (STAR Collaboration), these proceedings.
\bibitem{cite24}  D. Molnar and S.A. Voloshin, Phys. Rev. Lett. {\bf 91}, 92301 (2003).
\bibitem{cite25}  V. Greco, C.M. Ko and P. Levai, Phys. Rev. C {\bf 68}, 34904 (2003); R.J. Fries {\it et al.}, Phys. Rev. C {\bf 68}, 44902 (2003); R.C. Hwa and C.B. Yang, Phys. Rev. C {\bf 70}, 024905 (2004); R.J. Fries, S.A. Bass, and B. Muller, Phys. Rev. Lett. {\bf 94}, 122301 (2005).
\bibitem{cite26}  J. Adams {\it et al.} (STAR Collaboration), Phys. Rev. Lett. {\bf 92}, 112301 (2004).
\bibitem{cite27}  J. Adams {\it et al.} (STAR Collaboration), nucl-ex/0308033.
\bibitem{cite28}  L. Molnar {\it et al.} (STAR Collaboration), nucl-ex/0507027.
\bibitem{cite29}  J. Speltz {\it et al.} (STAR Collaboration), poster presented at this conference.
\bibitem{cite30}  P. Braun-Munzinger, I. Heppe, and J. Stachel, Phys. Lett. B {\bf 465}, 15 (1999); N. Xu and M. Kaneta, Nucl. Phys. {\bf A698}, 306 (2002).
\bibitem{cite31}  E. Schnedermann, J. Sollfrank, and U.W. Heinz, Phys. Rev. C {\bf 48}, 2462 (1993); U.A. Wiedemann and U.W. Heinz, Phys. Rev. C {\bf 56}, 3265 (1997).
\bibitem{cite32}  It is worth of noting that similar $T_{\rm CH}$ is also found in $pp$ and $e^+e^-$ collisions, which signifies the statistical nature of particle production (or hadronization). The extracted $\langle\beta_{\perp}\rangle$ is not zero in $pp$ which may be due to contamination of soft jet fragments in the measured $p_{\perp}$ spectra.
\bibitem{cite33}  S. Salur {\it et al.} (STAR Collaboration), these proceedings.
\end{thebibliography}
\end{document}